\documentclass[8pt]{article}

\usepackage{cite}

\usepackage{graphicx}
\graphicspath{%
    {converted_graphics/}
    {/}
}
\begin{document}

\title{Regular Reissner-Nordstr\"{o}m black hole solutions from linear electrodynamics}
\author{J. Ponce de Leon\thanks{E-mail:   	jpdel1@hotmail.com}\\ Laboratory of Theoretical Physics, Department of Physics\\ 
University of Puerto Rico, P.O. Box 23343, San Juan, \\ PR 00931, USA}

\maketitle

\begin{abstract}

In recent years there  have appeared in the literature a large number of static, spherically symmetric  metrics,  which are regular at the origin,  asymptotically flat,  and  have both an event  and a Cauchy horizon for certain range of the parameters. They have been interpreted as regular  black hole (BH) spacetimes,  and their physical source  attributed to   electric or magnetic
monopoles in a suitable chosen nonlinear electrodynamics. Here we show that these metrics can also be interpreted as exact solutions of the Einstein equations coupled to ordinary linear electromagnetism{\textemdash}i.e., as sources of the Reissner-Nordstr\"{o}m (RN) spacetime{\textemdash}provided the components of the effective energy-momentum tensor  satisfy the dominant energy condition (DEC). We use some well-known regular BH metrics to construct nonsingular RN black holes, where the singularity at the RN center  is replaced by a regular perfect fluid charged sphere (whose charge-to-mass ratio is not greater than $1$) which is   inside the RN inner horizon.

\end{abstract}

\medskip



\newpage

\section{Introduction}

In general relativity spacetime singularities have been present from the beginning, starting with the first solutions of Einstein's equations such as the Schwarzschild solution and the Friedmann solution. At first it was believed that these were mathematical artifacts induced by the requirement of 
 spherical
symmetry and the simplifying assumptions invoked to obtain the solutions. Today it is generally accepted{\textemdash}due to the famous singularity 
theorems (see, e.g., \cite{Hawking}){\textemdash}that    spacetime singularities are an inevitable feature for most
of  the  ``physically  reasonable"  models  of  Universe  and  gravitational  systems  within  the  framework  of  the
Einstein theory of gravity (for a review see, e.g., \cite{Joshi}).

However, the presence of singularities is usually regarded  as indicating the breakdown of the classical theory, requiring modifications  in  the  regions  where  the  spacetime
curvature becomes sufficiently high. The common opinion is that the problem of singularities could be solved by a consistent quantum theory of gravity. In the absence of such a theory, the issue of the resolution of singularities as produced by classical gravity remains open.

The validity of singularity theorems is  established under the hypothesis that certain conditions, which  can be roughly interpreted  as causality and energy conditions,  are met. If we leave aside the conditions that require causality, then the only possibility   to avoid singularities is a violation of energy conditions. In this regard, one way to eliminate the singularity during gravitational collapse was proposed by  Gliner \cite{Gliner}. He suggested that at very high densities, below some length scale, matter somehow makes a transition into a vacuumlike state, leading to the formation of a central core with de Sitter geometry \cite{falsevacuum}. 
This  hypothetical transition
avoids the conclusion of the Geroch-Hawking-Penrose theorems by violating the assumption
that matter obeys the strong energy condition (SEC). \footnote{Throughout the paper, we use a number of acronyms, e.g., BH (black hole), EM (Einstein-Maxwell), RN (Reissner-Nordstr\"{o}m),  EMT (energy-momentum tensor ), DEC  (dominant energy condition ), SEC (strong  energy condition). However, we avoid the use of  expressions like ``the EMT satisfies  the DEC but not the SEC,"  which make the paper difficult to read.}

Two years after Gliner's proposal, Bardeen \cite{Bardeen} presented  a metric  for an asymptotically flat, static, spherically symmetric spacetime  with a central de Sitter  core, which{\textemdash}at that time{\textemdash}served to establish the conditions for the existence of singularities.  Henceforth, that metric has been interpreted  as describing a regular  black hole (BH), since it is regular everywhere and has an inner and outer Killing horizon for certain values of the parameters.
From then on a number of models for regular BHs  have been proposed, which are   
 exact solutions to the Einstein field equations with different physical sources: (a) anisotropic fluids \cite{Dymnikova1, Mars, Borde, Hayward, Mbonye, Ansoldi}; (b) nonlinear electrodynamics \cite{Beato1, Beato2, Beato3, Beato4, Beato5, Bronnikov1, Dymnikova2, FanWang, Fernando}; (c)  scalar fields \cite{Bogojevic, Bronnikov2,  Bronnikov3}; and (d) modified gravity  \cite{Frolov, Berej, Zanchin1}. As expected, the discussion has been expanded to include rotating regular BHs \cite{Bambi, Gosh, Dymnikova3, Shadow1}.

After reviewing the literature, it is rather curious to note that ordinary linear electrodynamics is seldom considered for seeking regular BHs, 
with the notable exceptions of  the models having    a de Sitter interior which is joined  to the exterior Reissner-Nordstr\"{o}m (RN) field through a charged  shell (see e.g.,  \cite{Lemos1, Nami}), and the ones made out of charged phantom matter \cite{Lemos2}.
In this work we ask whether the BH metrics sourced by nonlinear electrodynamics can be used to construct regular RN black holes within the context of  linear electrodynamics. We will see that the answer to this question is positive if the components of the effective energy-momentum tensor  (EMT) satisfy the dominant energy condition (DEC).  We use some well-known regular BH metrics to construct nonsingular RN black holes, where the singularity at the RN center  is replaced by a regular perfect fluid charged sphere (whose charge-to-mass ratio is not greater than $1$) which is   inside the RN inner horizon.

The paper is organized as follows. In section $2$ we (i) introduce the notation and present the relevant field equations, (ii) discuss the conditions under which a static, spherically symmetric solution  of Einstein's equations coupled to nonlinear electrodynamics can be interpreted as a perfect-fluid solution  of Einstein's equations coupled to ordinary linear electromagnetism, and (iii) develop the boundary conditions. In section $3$ we consider the general class of nonsingular BHs recently considered by Fan and Wang \cite{FanWang} and construct the appropriate solutions to the ordinary Einstein-Maxwell (EM) equations, we find that for a range of parameters, such solutions represent BHs with the central singularity replaced by a charged perfect fluid sphere located inside the RN horizon. Similar results are obtained when  we extend the discussion to include the celebrated Dymnikova's vacuum nonsingular BH \cite{Dymnikova1}. Finally, in section $4$ we give a summary of the paper.

\section{Field equations}

Throughout  this work we use relativistic units where $c = G = 1$ and the sign conventions are those of Landau and Lifshitz \cite{Landau}.
For nonlinear electrodynamics in general relativity we consider the action 

\begin{equation}
\label{action}
S = - \frac{1}{16\, \pi}\, \int{\sqrt{- g}\, [R + L(F)]\, d^4 x},
\end{equation}
where $R$ is the scalar curvature,   $L$ is a function of $F = F_{\alpha \beta}\, F^{\alpha \beta}$, and  $F_{\mu \nu} = \partial_{\mu} A_{\nu} - \partial_{\nu} A_{\mu}$ is the electromagnetic field tensor. The Lagrangian density of the electromagnetic field    is $\Lambda = - \frac{1}{16\, \pi}\, L(F)$. Thus, the field equations for gravity are

\begin{equation}
\label{equations for gravity}
G_{\lambda \rho} = R_{\lambda \rho} - \frac{1}{2}\, g_{\lambda \rho}\, R = 8\, \pi\, T_{\lambda \rho},
\end{equation}
where $T_{\lambda \rho}$ represents the EMT associated with the nonlinear electromagnetic field, which is  
\begin{equation}
\label{ETM for nonlinear electrodynamics}
T_{\lambda \rho} = \frac{1}{4\,\pi}\, \left(\frac{L}{4}\, g_{\lambda \rho} - L_{\!_{F}}\, F_{\mu \lambda}\, F_{\nu \rho}\,g^{\mu \nu}\right),\;\;\;\;L_{\!_{F}} \equiv \frac{d L}{d F}.
\end{equation}
The tensor $F^{\mu \nu}$ is governed by the equations 
\begin{eqnarray}
\left(L_{\!_{F}}\, F^{\mu \nu}\right)_{; \mu} &=& 0,\\
F_{\mu \nu; \lambda} + F_{\nu \lambda; \mu} + F_{\lambda \mu; \nu} &=& 0\label{Bianchi identity fro the EM field}.
\end{eqnarray}
In the case where $L = F$ as well as  in the Maxwell weak-field limit, when $L(F) \stackrel{F \rightarrow 0}{\longrightarrow}  F$, the above equations reduce to the familiar set of (linear) EM equations, as expected.

The most general static spherically symmetric metric can be written as

\begin{equation}
\label{The most general static spherically symmetric metric}
d s^2 = f(u)\, d t^2 - h(u)\, d u^2 - r^2(u)\,d \Omega^2,
\end{equation}
where   $d\, \Omega^2 = (d \theta^2 + \sin^2 \theta\, d \phi^2)$ and $u$ is a radial coordinate. 
Due to the spherical symmetry, the EMT for an arbitrary kind of matter can be written as

\begin{equation}
\label{EMT 2}
T_{\mu}^{\mu} = \mbox{diag}\,(\epsilon, \, - p_{r}, \, - p_{\perp}, \, - p_{\perp}),
\end{equation}
where the energy density $\epsilon$, the radial pressure $p_{r}$ and the transverse pressure $p_{\perp}$ are functions of $u$. A physically reasonable EMT must  be free of singularities, have nonnegative energy density and satisfy the  local conservation of stress-energy $\nabla_{\mu}\, T^{\mu}_{\nu} = 0$. In addition, ordinary/baryonic matter is expected to obey the energy conditions.\footnote{For the EMT (\ref{EMT 2}) the dominant energy condition (DEC) requires $\epsilon \geq |p_{r}|$, $\epsilon \geq |p_{\perp}|$; the weak energy condition (WEC) requires  $\epsilon > 0$, $\epsilon + p_{r} \geq  0$, $\epsilon + p_{\perp} \geq 0$; the null energy
condition (NEC) requires $\epsilon + p_{r} \geq 0$, $\epsilon + p_{\perp} \geq 0$; the strong energy condition (SEC) requires
$\epsilon + p_{r} + 2\, p_{\perp} \geq 0$, $\epsilon + p_{r} \geq 0$, $\epsilon + p_{\perp} \geq 0$. These are not mutually independent; if the DEC is satisfied, then the weak and the null  energy conditions are automatically satisfied as well. Also, the NEC is implied by the strong energy condition.}

By virtue of the spherical symmetry the only nonvanishing components of $F_{\mu \nu}$ are $F_{0 1} = - F_{1 0}$ and $F_{2 3} = - F_{3 2}$. If we set 

\begin{equation}
\label{field in terms of F}
F_{0 1}\, F^{0 1} = - E^2, \;\;\;\;\;F_{2 3}\, F^{2 3} = B^2, 
\end{equation}
where $E$ and $B$ only depend on the radial coordinate, then

\begin{eqnarray}
F &=& 2\, (B^2 - E^2)\label{explicit form of F},\\
T^{0}_{0} = T_{1}^{1} &=& \frac{1}{4\, \pi}\, \left(\frac{L}{4} + L_{\!_{F}}\, E^2\right)\label{T00 = T11, general case},\\
T^{2}_{2} = T_{3}^{3} &=& \frac{1}{4\, \pi}\, \left(\frac{L}{4} -  L_{\!_{F}}\, B^2\right)\label{T22 = T33, general case}.
\end{eqnarray}
As a consequence of (\ref{T00 = T11, general case}), the equation $G_{0}^{0} = G_{1}^{1}${\textemdash}evaluated for the line element (\ref{The most general static spherically symmetric metric}){\textemdash}can be easily integrated to obtain

\begin{equation}
h(u) = \frac{\mbox{constant}}{f(u)}\, \left(\frac{d r}{d u}\right)^2.
\end{equation}
Substituting this into (\ref{The most general static spherically symmetric metric}), setting the constant of integration equal to $1$, and using $r = r(u)$ as the new radial coordinate we arrive at the simplified line element

\begin{equation}
\label{metric in temrs of f}
d s^2 = f(r)\, d t^2 - \frac{d r^2}{f(r)} - r^2\,(d \theta^2 + \sin^2 \theta\, d \phi^2).
\end{equation}
In these coordinates the nonvanishing components of the generic EMT (\ref{EMT 2}) are

\begin{eqnarray}
\epsilon = - p_{r} &=& \frac{1}{8\,\pi\,r^2}\left[1 - (r\, f)'\right]\label{FE for epsilon, Dimnikova},\\
p_{\perp} =   &=& \frac{(r^2\,f')'}{16\, \pi\, r^2} \label{transversal pressure},
\end{eqnarray}
where a prime denotes differentiation with respect to $r$. We note that

\begin{equation}
\label{epsilon decreases outward when DEC holds}
p_{\perp} = - \epsilon - \frac{r\, \epsilon'}{2},        
\end{equation}
which implies that (i) $p_{\perp} \approx - \epsilon$ near the (regular) center, and 
(ii) $\epsilon' < 0 \Leftrightarrow (\epsilon + p_{\perp}) > 0$. 
Thus, when $\epsilon > 0$ and $\epsilon' < 0$ the weak and null energy conditions are automatically satisfied. 
The SEC is violated in the central region, because it stipulates that  $p_{\perp} \geq 0$ and $\epsilon + p_{\perp} \geq 0$ when $p_{r} = - \epsilon$. The DEC, which now reduces  to $\epsilon \geq |p_{\perp}|$, is not necessarily satisfied.

In terms of the mass function

\begin{equation}
m(r) = 4\, \pi \int_{0}^{r}{\bar{r}^2\, \epsilon(\bar{r})\, d \bar{r}},
\end{equation}
the field equation (\ref{FE for epsilon, Dimnikova}) can be integrated as

\begin{equation}
\label{f in terms of m(r)}
f(r) = 1 - \frac{2\, m(r)}{r},
\end{equation}
where, to avoid a singularity at the origin,  the constant of integration has been set equal to zero.
Equivalently, the mass function can be written as

\begin{equation}
\label{mass function}
m(r)  = \frac{r}{2}\, (1 - f).
\end{equation}
However, the active gravitational mass inside a  volume $V$ is given by the  Tolman-Whittaker (TW) formula \cite{Landau}

\begin{equation}
\label{Tolman-Wittaker mass}
M = \int {(T_{0}^{0} - T_{1}^{1} - T_{2}^{2} - T_{3}^{3})\, \sqrt{- g}\, d V},
\end{equation}
which in the case under consideration reduces to

\begin{equation}
\label{simplified form of TW mass}
M(r) = \frac{r^2\, f'}{2}.
\end{equation}
Clearly the TW mass calculated between the center and any point inside (outside) the
first horizon (event horizon) is negative (positive). 
In the region where $f < 0$  the coordinate $r$ is timelike ant $t$ is spacelike. Therefore, (\ref{Tolman-Wittaker mass}) in that region  no longer has the  direct physical meaning of active gravitational mass \cite{Zaslavskii}.

\subsection{Charged perfect fluid interpretation}

In general relativity the same geometry can be engendered by different material distributions. For example,  under some circumstances the EMT of a generic anisotropic fluid (\ref{EMT 2})  can be represented as a multicomponent fluid. Recently, it has been shown that static spheres of anisotropic fluid  can be represented as a  linear combination of  perfect fluid, electromagnetic field, and minimally coupled scalar field \cite{Visser}. Typically the decomposition procedure is not unique because  the number of independent functions in  the multicomponent  model is greater than the one in the anisotropic one-fluid model.
Below we will see that the matter distribution supporting the line element (\ref{metric in temrs of f}) can be interpreted as a charged perfect fluid, with good physical properties,  if  the DEC ($\epsilon \geq |p_{\perp}|$) holds. 

Thus, we set 

\begin{equation}
\label{superposition of EMT's}
T_{\mu \nu} = \tau_{\mu \nu} + {\cal{E}}_{\mu \nu},
\end{equation}
where $\tau_{\mu \nu}$ and ${\cal{E}}_{\mu \nu}$  represent the EMT for  perfect fluid and   linear ($L = F$) electromagnetism, respectively. Namely, 

\begin{eqnarray}
\tau_{\mu \nu} &=& (\rho + p)\,u_{\mu}\, u_{\nu} - p\, g_{\mu \nu},\\
{\cal{E}}_{\mu \nu} &=& \frac{1}{4\, \pi}\, \left(- F_{\mu \lambda}\, F_{\nu \sigma}\, g^{\lambda \sigma} + \frac{1}{4}\, g_{\mu \nu}\, F_{\alpha \beta}\, F^{\alpha \beta}\right)\label{cal E}.
\end{eqnarray}
Here $u^{\nu}$, $\rho$ and $p$ are  the four-velocity, energy density and isotropic pressure of the fluid, respectively.
From (\ref{cal E})  we find

\begin{equation}
{\cal{E}}^{\mu \nu}_{\;\;\;\;; \nu} = - F^{\mu \lambda} J_{\lambda},
\end{equation}
where the four-vector $J^{\mu}$ is defined by the equation

\begin{equation}
\label{definition of J}
F^{\mu \nu}_{\;\;\;\;; \nu} = \frac{1}{\sqrt{- g}} \, \frac{\partial (\sqrt{- g}\, F^{\mu \nu})}{\partial x^{\nu}} =  - 4\, \pi\, J^{\mu}.
\end{equation}
Since $J^{\mu}_{\;\;; \mu} = 0$, it can be interpreted as an (effective) current density four-vector. Thus,  (\ref{definition of J}) is equivalent to  the second pair of Maxwell equations \cite{Landau}. Consequently, the conservation equation $T^{\mu \nu}_{\;\;\;\;; \nu} = 0$ can be written as

\begin{equation}
\label{conservation equation}
\tau^{\mu \nu}_{\;\;\;\;;\nu} = F^{\mu \nu}\, J_{\nu}.
\end{equation}
Finally, the proper electrical charge density ${\bar{\rho}}_{e}$ is introduced through the relation

\begin{equation}
\label{proper charge density}
J^{\mu} = {\bar{\rho}}_{e}\, u^{\mu}.
\end{equation}
Thus, by definition ${\bar{\rho}}_{e}^2 = (J_{\mu} J^{\mu})$.

Let us now go back to the case of a spherical  distribution of matter.  The EMT (\ref{superposition of EMT's}) in the comoving frame, where $u^{\mu} = (\delta^{\mu}_{0}/\sqrt{g_{00}})$,    
 reduces to 

\begin{equation}
\label{EMT for charged perfect fluid 2}
T_{\mu}^{\mu} = \mbox{diag}\,(\rho + W, \, - p + W, \, - p - W, \, - p - W),
\end{equation}
with 
\begin{equation}
\label{energy density of the electromagnetic field}
W  = \frac{E^2 + B^2}{8\, \pi}, 
\end{equation}
which is the energy density of the electromagnetic field.
Equating term by term the components of 
tensors (\ref{EMT 2}) and (\ref{EMT for charged perfect fluid 2}) we obtain a system of three equations in three unknowns from which we get

\begin{eqnarray}
\rho &=& \epsilon - \frac{1}{2}\, (p_{\perp} - p_{r}),\label{solution for rho}\\
p &=& \frac{1}{2}\, (p_{r} + p_{\perp}),\label{solution for p} \\
 W &=& \frac{1}{2}\, (p_{\perp} - p_{r})\label{solution for E squared}.
\end{eqnarray}
In the case under consideration $p_{r} = - \epsilon$,  as a result of (\ref{EMT 2})  and (\ref{T00 = T11, general case}).  Therefore,  
   the above  equations  reduce to

\begin{eqnarray}
\rho = - p &=& \frac{1}{2}\,(\epsilon - p_{\perp}),\label{rho, general expression}\\
W &=& \frac{1}{2}\, (\epsilon + p_{\perp}).\label{W, general expression}
\end{eqnarray}
Now we can show that in the present interpretation there is no room for linear EM magnetic monopoles, even if the original spacetime (\ref{metric in temrs of f}) is attributed to a magnetic monopole in a nonlinear electromagnetic theory. In fact, from (\ref{field in terms of F}) we obtain $F_{2 3} = \pm \, r^2\, \sin \theta\, B(r)$. Substituting this into (\ref{Bianchi identity fro the EM field}) -- with $\mu = 1$, $\nu = 2$, $\lambda = 3$ -- we get $B(r) = Q_{m}/r^2$, where $Q_{m}$ is a constant of integration. Then,  from  (\ref{energy density of the electromagnetic field}) and  (\ref{W, general expression}) it follows that $E^2 \stackrel{r \rightarrow 0}{\longrightarrow} \;  - Q_{m}^2/r^4$ 
for any spacetime satisfying regularity conditions at the origin. To avoid  this  unwanted consequence  in what follows we  set $Q_{m } = 0$.  Thus, 

\begin{equation}
\label{E, general expression}
E^2  = 4\, \pi\, (\epsilon + p_{\perp}).
\end{equation}
According to (\ref{epsilon decreases outward when DEC holds}), at the center $\rho = \epsilon$ and $E = 0$. The DEC ($\epsilon \geq |p_{\perp}|$) ensures that both $\rho$ and $E^2$ are nonnegative.
The first zero of the equation $\epsilon = p_{\perp}$
(the closest to the center, say $r = r_{s}$) represents the boundary surface of the distribution since both  the matter density $\rho$ and pressure $p$ vanish there. Then   (\ref{metric in temrs of f})  can be used to represent the interior of a charged perfect fluid sphere of coordinate radius $r = r_{s}$. Otherwise, if $\epsilon > |p_{\perp}|$ holds everywhere, the  charged perfect fluid occupies the whole space. 

In the static case, from (\ref{definition of J}) we get 

\begin{equation}
\label{electric field}
E(r) = \frac{4\, \pi}{r^2}\, \int_{0}^{r} \bar{r}^2\, \rho_{e}(\bar{r})\, d \bar{r} \equiv \frac{q(r)}{r^2},
\end{equation}
where $E(r) = - \sqrt{- g_{0 0}\, g_{1 1}}\, F^{0 1}$ is the usual radial electric field\footnote{From (\ref{field in terms of F})  it follows that  $E(r) = \pm \sqrt{- g_{0 0}\, g_{1 1}}\, F^{0 1}$, we choose the negative sign in order to get the familiar equations in electrodynamics.} intensity, 
 $q(r)$ is the electric charge inside a sphere of coordinate radius $r$ and $\rho_{e}$ is the charge density which is related to the proper charge density $\bar{\rho}_{e}$ by

\begin{equation}
\rho_{e} =  {\sqrt{- g_{1 1}}}\, {\bar{\rho}_{e}}.
\end{equation}
Similarly, (\ref{conservation equation}) reduces to the Tolman-Oppenheimer-Volkoff (TOV) equation of hydrostatic equilibrium for a charged perfect fluid sphere, viz.,

\begin{equation}
\label{TOV}
p' + \left(\rho + p \right)\, (\ln \sqrt{g_{00}})' = E\,\rho_{e}. 
\end{equation}
For the models under consideration $p = - \rho$. Therefore this equation reduces to $p' = E\,\rho_{e}$. This means that  
the pressure gradient -- which exerts a force towards the center -- is only balanced by the electrostatic repulsion.

Finally, the components of the EMT (\ref{EMT for charged perfect fluid 2}) in terms of the metric (\ref{metric in temrs of f}) are given by

\begin{eqnarray}
\rho = - p &=& \frac{1}{16\, \pi\, r^2}\, \left(1 - f - 2\, r\, f' - \frac{r^2\, f''}{2}\right)\label{rho and p in terms of f}, \\
E^2 &=& \frac{1}{2\, r^2}\,\left(1 - f + \frac{r^2\, f''}{2}\right)\label{E in terms of f}.
\end{eqnarray}

\subsection{Boundary conditions}

The solution of the EM equations for $r > r_{s}$, outside the sphere,  is given
 by the RN field which in curvature coordinates has the form

\begin{equation}
\label{RN metric 2}
d s^2 = f_{\!_{RN}}(r)\, d t^2 -  \frac{d  r^2}{f_{\!_{RN}}(r)} - r^2 \, \left(d \theta^2  + \sin^2 \theta d \phi^2\right),
\end{equation}
with

\begin{equation}
\label{f for RN}
f_{\!_{RN}}(r) = 1 - \frac{2\, M}{r} + \frac{Q^2}{r^2}.
\end{equation}
The radial electric field is

\begin{equation}
\label{radial external electric field 2}
E(r) = \frac{Q}{r^2},
\end{equation}
In the above expressions $M$ and $Q$ are the total mass and charge, respectively, which are related to the parameters  of the internal solution through the boundary conditions. 
Later we will need the expressions for the mass function and the TW mass in the external region{\textemdash}say $m_{\!_{RN}}(r)$ and $M_{\!_{RN}}(r)$, respectively. They are given by

\begin{eqnarray}
m_{\!_{RN}}(r) &=& M - \frac{Q^2}{2\, r},\label{mass function for RN} \\
M_{\!_{RN}}(r) &=& M - \frac{Q^2}{r}.\label{TW mass for RN}
\end{eqnarray}
These equations show that both $m_{\!_{RN}}(r)$ and $M_{\!_{RN}}(r)$ become negative when $r$ is sufficiently small.

To match the internal and external solutions
across  the boundary $r = r_{s}$, we require continuity of the first and second fundamental forms. For this,  we need continuity of $f(r)$ and  $T_{1}^{1}$, respectively. Since $p(r_{s}) = 0$, in the absence of surface concentration of charge the latter condition demands continuity of $E(r)$, which in turn{\textemdash}by virtue of  (\ref{rho and p in terms of f}) and (\ref{E in terms of f}){\textemdash}requires continuity of $f'$.
In summary, at the boundary we require

\begin{equation}
f_{s} = f(r_{s}) = f_{\!_{RN}}(r_{s}), \;\;\;\;f'_{s} = f'(r_{s}) = f'_{\!_{RN}}(r_{s}).
\end{equation}
Using   (\ref{f for RN}), these expressions constitute a system of two algebraic equations which allow us to obtain $M$ and $Q$ in terms of $f_{s}$ and $f'_{s}$. The solution is

\begin{eqnarray}
M &=& r_{s}\, (1 - f_{s} - \frac{r_{s}\, f'_{s}}{2}),\label{BC 1}\\
Q^2 &=& r_{s}^2\, (1 - f_{s} - r_{s}\, f'_{s}).\label{BC 2}
\end{eqnarray}
These expressions guarantee the continuity of the mass function [Eqs. (\ref{mass function}) and (\ref{mass function for RN})] and TW mass [Eqs. (\ref{simplified form of TW mass}) and (\ref{TW mass for RN})] across the surface of the sphere.

{\it Restrictions on $r_{s}$:} As a consequence of the DEC the possible values of $r_{s}$ are bounded below by 

\begin{equation}
\label{rmin}
r_{min} = \frac{2\, Q^2}{3\, M}.
\end{equation}
In fact, from (\ref{epsilon decreases outward when DEC holds}) it follows that $\epsilon' \leq 0$ if $\epsilon \geq - p_{\perp}$, which means $\epsilon(r) \geq \epsilon(r_{s})$. Combining this with $f_{s} = f_{\!_{RN}}(r_{s})$ we obtain

\[
\frac{8\, \pi}{r_{s}}\,\int_{0}^{r_{s}} r^2 \left(\rho + \frac{E^2}{8\, \pi}\right)\, dr = \frac{2\, M}{r_{s}} - \frac{Q^2}{r_{s}^2} \geq \frac{Q^2}{3\, r_{s}^2}.
\]
Solving the inequality for $r_{s}$ we obtain $r_{s} \geq r_{min}$, given by (\ref{rmin}).  This guarantees that $m_{\!_{RN}}(r)$ is positive for all values of $r_{s}$, while $M_{\!_{RN}}(r)$ is negative inside and in the vicinity of the sphere when $r_{min} \leq r_{s} < 3\, r_{min}/2$.

In addition, the charged spheres can be inside the RN horizon if

\begin{equation}
\label{values of alpha for which rs less than r-}
\alpha \equiv \frac{|Q|}{M} \in \, (\frac{\sqrt{3}}{2}, \, 1]\,  \approx (0.866, \, 1].
\end{equation}
Indeed,  when  $|Q| = M$ the Killing horizon is located at $r = r_{\ast} = M$, so that $r_{min} = \frac{2}{3}\, r_{\ast}$. 
Similarly, when $|Q| < M$ the inner horizon is located at $r = r_{-} = M \, [1 -\sqrt{1 - (Q/M)^2}]$. Consequently, $r_{min} < r_{-}$ when (\ref{values of alpha for which rs less than r-}) is satisfied.

\section{Regular RN black holes }

In this section,  we apply the above equations to  some well-known regular BH spacetimes and show that  they can be interpreted as exact solutions to the EM equations describing  spherical distributions of charged perfect fluid. The coordinate radius of the spheres is determined by the equation $\epsilon =  p_{\perp}$. They have several interesting properties: (i) the charge-to-mass ratio $\alpha = |Q|/M$ is bounded below; (ii) the spheres having  $\alpha \leq 1$ are hidden behind the RN horizons and are  gravitationally repulsive inside as well as in their  vicinity, although the repulsive region is covered by the event horizon; (iii)  when $\alpha > 1$ the vicinity of the spheres can be gravitationally attractive or repulsive, depending on the model; (iv) the energy density at the (regular) center is proportional to $\alpha^{- 6}\, M^{- 2}$, for an object one  solar mass it is  about $10^{20}\, kg/m^3${\textemdash}although an increase (decrease) in $M$ leads to a decrease (increase)  of this quantity.

\subsection{General class of nonsingular RN black holes}

First we consider the line element (\ref{metric in temrs of f}) generated by the  metric function 
\begin{equation}
\label{general function}
f(r) = 1 - \frac{2\, m\, r^{\sigma - 1}}{\left(r^{\beta} + K\right)^{\sigma/\beta}},
\end{equation}
where $\sigma > 1$, $\beta \ > 0$, $K \geq 0$ and $m \geq 0$ are constants. It reduces to the Schwarzschild vacuum solution with mass $m$ for $K = 0$. This line element has recently been discussed in the context of nonlinear electrodynamics by Fang and Wang \cite{FanWang}. It includes the Bardeen solution \cite {Bardeen} for $\sigma = 3$ and $\beta = 2$,  the Hayward solution \cite{Hayward} for $\sigma = \beta = 3$, as well as a new class  of solutions considered by Fan and Wang for $\beta = 1$. For certain values of the parameters these can be interpreted  as asymptotically flat BHs.\footnote{A horizon corresponds to a zero of $f(r)$; the outermost zero is the event horizon of the BH. The  function (\ref{general function}) has a minimum at $r = r_{min} = [K\, (\sigma - 1)]^{1/\beta}$, viz.,

\[
f_{min} = f(r_{min}) = 1 - \left(\frac{K_{crit}}{K}\right)^{1/\beta},
\]
where

\[
K_{crit} = \frac{(2\, m)^{\beta}}{(\sigma - 1)}\, \left(\frac{\sigma - 1}{\sigma}\right)^{\sigma}.
\]
Thus, (\ref{general function}) has no zeros if $K > K_{crit}$, one double zero at $r = r_{\ast} = 2\, m\, \left(\frac{\sigma - 1}{\sigma}\right)^{\sigma/\beta}$ if $K = K_{crit}$, and two simple zeros if $K < K_{crit}$.}

Our aim is to show that (\ref{general function}) can be used to represent perfect fluid charged spheres in ordinary EM theory. To begin with  we use (\ref{FE for epsilon, Dimnikova}) and (\ref{transversal pressure}) to evaluate the effective EMT. We obtain

\begin{eqnarray}
\epsilon = - p_{r} &=& \frac{\sigma\, m\, K\, r^{\sigma - 3}}{4\, \pi\, \left(r^{\beta} + K\right)^{(\sigma + \beta)/\beta}}\label{epsilon for general class},\\
p_{\perp} &=& \frac{\epsilon \, [(1 - \sigma)\, K + (1 + \beta)\, r^{\beta}]}{2\, (r^{\beta} + K)}.
\end{eqnarray}

The parameter $\sigma$ is specified by the equation of state $(p_{\perp}/\epsilon)$ near the origin as 
\begin{equation}
\label{determination of sigma}
\sigma = 1 - 2\, \left.\frac{p_{\perp}}{\epsilon}\right|_{r = 0}.
\end{equation}
If $\left.(p_{\perp}/\epsilon)\right|_{r = 0} \in (0, \, - 1]$, then $\sigma \in (1, \, 3]$. 
It determines the behavior of the energy density: when $\sigma \leq 3$, $\epsilon$  is positive everywhere and monotonically decreases outward; when $\sigma > 3$, $\epsilon$ increases from zero at the origin up to a maximum value and then decreases to zero for large values of $r$. When $\sigma = 3$ the central region is de Sitter-like with $\epsilon = - p_{r} \approx - p_{\perp} = \frac{3\, m}{4\, \pi}\,K^{- 3/\beta}$.

\medskip

The DEC is satisfied in different regions depending on the choice of $\beta$. Namely,

\begin{eqnarray}
\beta \leq &1&: r^{\beta} \geq \frac{K\, (\sigma - 3)}{3 + \beta}, \label{DEC for beta leq than 1}\\
\beta > &1&:\frac{K\, (\sigma - 3)}{3 + \beta} \leq r^{\beta} \leq \frac{K\, (1 + \sigma)}{\beta - 1}.\label{intermediate region}
\end{eqnarray}
Thus, $\epsilon \geq |p_{\perp}|$ everywhere only if $\sigma \leq 3$ and $\beta \leq 1$. Otherwise it only holds in the  region (\ref{intermediate region}).

Consequently, the matter distribution supporting
the metric (\ref{metric in temrs of f})  with the function $f(r)$ given by (\ref{general function}) can be interpreted as a charged perfect fluid when  $\sigma \leq 3$, which includes the origin for any  $\beta > 0$. The corresponding energy density, isotropic pressure, and electric field are

\begin{eqnarray}
\rho =  - p &=& \frac{\epsilon \, [K\,(\sigma + 1) - (\beta - 1)\, r^{\beta}]}{4\, (r^{\beta} + K)},\label{rho and p for the general class} \\
E^2 &=& \frac{2\,\pi\,\epsilon\, [(3 - \sigma)\, K + (3 + \beta)\, r^{\beta}]}{(r^{\beta} + K)}. \label{E for the case with K, sigma , beta }
\end{eqnarray}

\paragraph{Boundless charged configuration:} When $0 <\beta \leq 1$ and $1 < \sigma \leq 3$ the DEC (\ref{DEC for beta leq than 1}) is satisfied everywhere and  the pressure (\ref{rho and p for the general class}) never vanishes. Therefore, (\ref{rho and p for the general class}){\textendash}(\ref{E for the case with K, sigma , beta }) can be interpreted as an unbounded spherical distribution of charged perfect fluid. However, the total charge $|Q| = \left[r^2\, E(r)\right]_{r \rightarrow \infty}$ is finite{\textemdash}and equal to $\sqrt{2\, K \sigma\,  m}${\textemdash}only when\footnote{For $0 < \beta < 1$ the total electric charge diverges because $(r^4\, E^2) \stackrel{r \rightarrow \infty}{\longrightarrow}\, \frac{m\, \sigma\, K\,(3 + \beta)}{2}\, r^{1 - \beta}$.} $\beta = 1$. As expected, this is consistent with the limiting expression

\[
f_{\!_{\beta = 1}}(r) \stackrel{r \rightarrow \infty}{\longrightarrow}\, 1 - \frac{2\,m}{r} + \frac{2\, K\, \sigma\, m}{r^2} - {\cal{O}}\left(\frac{1}{r^3}\right),
\]
which also shows that $m$ is the total mass of the configuration. We arrive at the same conclusion  by using the TW mass, viz.,  $M(r) \stackrel{r \rightarrow \infty}{\longrightarrow} M = m$.

For this case, the metric function (\ref{general function}) can be parametrized by mass and charge as

\begin{equation}
f(u) = 1 - \frac{2}{u}\, \left(1 + \frac{\alpha^2}{2\, \sigma\, u}\right)^{- \sigma},
\end{equation}
 where $u = \frac{r}{M}$,  and $0 \leq u < \infty$. The condition $f (u) > 0$ imposes a lower limit on $\alpha$, say $\alpha_{min}$, viz.,  

\begin{equation}
\label{lower limit of alpha, 1}
\alpha > \alpha_{min} = 2\,\left(\frac{\sigma - 1}{\sigma}\right)^{(\sigma - 1)/2}.
\end{equation}
By virtue of (\ref{determination of sigma}), $\alpha_{min}$ is determined by the equation of state at the center. Note that $\alpha_{min}$ is greater than $1$ for all values of $\sigma \in (1, \, 3]$.

When $\sigma = 3$, the density at the center is completely determined by  $Q$ and $M$. Indeed, 
for the case under consideration, when $\beta = 1$ (which implies $K = \frac{Q^2}{6\,M}$) from (\ref{epsilon for general class}) and (\ref{rho and p for the general class}),  we get

 \begin{equation}
\label{rho M squared for unbounded distribution}
\rho_{c} = \frac{162}{\alpha^6\, \pi \, M^2},
\end{equation}
where $\rho_{c}$ is the energy density at the center. In general, $\rho_{c}\, M^2$ is bounded above as  $\alpha$ is bounded below. For $\sigma = 3$, we find $\alpha_{min} = 4/3$ and thus $\rho_{c}\, M^2 < 9.178$, approximately. If we apply this inequality to an object of one solar mass ($M = M_{\odot}$) we find $\rho_{c} < \rho_{c_{max}} \approx 5.683 \times 10^{21} kg/m^3$.

\paragraph{Charged perfect fluid spheres:} When $\beta > 1$ and $1 < \sigma \leq 3$ the DEC (\ref{intermediate region}) is met from the origin up to a maximum value of $r$ which determines the boundary of the sphere,  where the density and pressure vanish.

Thus, from (\ref{rho and p for the general class}) $K$ can be expressed in terms of $r_{s}$, the coordinate radius of the sphere, as 

\begin{equation}
\label{K in terms of rs}
K = \frac{(\beta - 1)\, r_{s}^{\beta}}{1 + \sigma}.
\end{equation}
Using the boundary conditions (\ref{BC 1}) and (\ref{BC 2}) at $r = r_{s}$ we find

\begin{eqnarray}
r_{s} &=& \frac{M\, (1 + \sigma)\, \beta \, \alpha^2}{2\, \sigma\, (\beta - 1)},\label{radius of the sphere with beta, sigma} \\
m &=& \frac{M}{\beta}\, \left(\frac{\sigma + \beta}{1 + \sigma}\right)^{(\sigma + \beta)/\beta}\label{m of the sphere with beta, sigma}.
\end{eqnarray}
Note that $m$ is no longer the total mass of the charged configuration.

Taking these results into account, the metric function (\ref{general function}) can be written as

\begin{equation}
\label{final form of f for the sphere with beta, sigma}
f(x) = 1 - \frac{4\, \sigma\, (\beta - 1)\, \left(\sigma + \beta\right)^{(\sigma + \beta)/\beta}\, x^{\sigma - 1}}{\alpha^2 \, \beta^2\, (1 + \sigma)^2\, \left[(1 + \sigma)\, x^{\beta} + \beta - 1\right]^{\sigma/\beta}}, \;\;\;\;x = \frac{r}{r_{s}}
\end{equation}
This function must be positive in the whole range $x \in [0, \, 1]$. Leaving aside the details of the analysis, we find that
this condition  imposes a lower bound on $\alpha$, viz.,

\begin{equation}
\label{alpha min for the general class}
\alpha > \tilde{\alpha}_{min} =  \frac{2\, \sqrt{\sigma}}{1 + \sigma}. 
\end{equation}
Otherwise, when $\alpha \leq \tilde{\alpha}_{min}$ it   cannot be satisfied by any  $\beta$. Note that unlike the unbounded case (\ref{lower limit of alpha, 1}), now $\tilde{\alpha}_{min}$ is  less than $1$ for all $\sigma \in (1, \, 3]$. We also find that  $f(x) > 0$ is satisfied by {\it any}  $\beta > 1$ and $\sigma \in (1, \, 3]$ if

\begin{equation}
\label{alpha star}
\alpha \geq \alpha_{\star} = 2\, \left(\frac{\sigma - 1}{\sigma}\right)^{(\sigma - 1)/2}.
\end{equation}
Consequently,  the condition $f(x) > 0$ only restricts  the values of $\beta$ when $\tilde{\alpha}_{min} < \alpha < \alpha_{\star}$, e.g., $\sqrt{3}/2 < \alpha < 4/3$ for $\sigma = 3$. Parenthetically, we point out that   $\alpha_{\star}$ is equal to $\alpha_{min}$ introduced in (\ref{lower limit of alpha, 1}),   although they refer to different physical scenarios.   Obviously,  $\alpha_{\star} > \tilde{\alpha}_{min}$ for $\sigma \in (1, \, 3]$. 

In the range 
$\alpha \in (1, \, \alpha_{\star})${\textemdash}e.g., $\alpha \in (1, \, 4/3)$ for  
$\sigma = 3${\textemdash}the condition $f(x) > 0$ leads to a cumbersome transcendental inequality  between $\alpha$, $\sigma$,  and $\beta$, namely, 

\begin{equation}
\label{transcendental inequality}
1 < \alpha < \alpha_{\star}:\;\;\;\;   \alpha^2 - 4\, \beta^{- 2}\,\left[\sigma^{(\beta - \sigma)}\, (\beta - 1)^{(\beta -1)}\, (\sigma + \beta)^{(\sigma + \beta)}\, (\sigma - 1)^{(\sigma - 1)}\, (1 + \sigma)^{- (\sigma - 1 + 2\, \beta)}\right]^{1/\beta} > 0.
\end{equation}

However, the range of interest of $\alpha$ is $\alpha \in (\tilde{\alpha}_{min}, \, 1]$,  which allows the presence of RN black holes. In that range we find that $f(x) > 0$ only if
\begin{equation}
\tilde{\alpha}_{min} < \alpha \leq 1:\;\;\;\;      \label{solution that satisfies the above-mentioned condition}
\beta > \frac{2\,\sigma\,  [(\sigma - 1) + (\sigma + 1)\, \sqrt{1 - \alpha^2}]}{\alpha^2\, (1 + \sigma)^2 - 4\, \sigma},
\end{equation}
which is one of the solutions to the inequality $f(1) > 0$. Note that the denominator here is positive by virtue of (\ref{alpha min for the general class}). 
The last inequality  has several interesting consequences:

\medskip

$\bullet$ {\it{Static spheres hidden behind RN horizons:}} From (\ref{solution that satisfies the above-mentioned condition}) it follows that the charged spheres with $\alpha \leq 1$ are located inside the RN horizons. To illustrate this we consider the cases $\alpha = 1$ and $\alpha < 1$ separately.

\medskip

(i)  If $\alpha  = 1$, then from (\ref{solution that satisfies the above-mentioned condition}) we obtain $\beta > \frac{2\, \sigma}{\sigma - 1}$. It is easy to verify that this also follows from the inequality $r_{s} < r_{\ast} = M$, where $r_{s}$ is given by (\ref{radius of the sphere with beta, sigma}). Thus, the spheres with $\alpha = 1$ are inside the Killing horizon $r = r_{\ast} = M$. In fact, from (\ref{radius of the sphere with beta, sigma}){\textemdash}with $\alpha = 1$ and  $\frac{2\, \sigma}{\sigma - 1} < \beta < \infty${\textemdash}we get

\begin{equation}
\label{rs/rH for alpha = 1}
\frac{1 + \sigma}{2\, \sigma} < \left(\frac{r_{s}}{r_{\ast}}\right) < 1
\end{equation}

\medskip

(ii) If $\alpha < 1$, then (\ref{solution that satisfies the above-mentioned condition})  is also the solution to the inequality

\begin{equation}
\label{rs less than inner horizon}
r_{s} < r_{-} = M\,(1 - \sqrt{1 - \alpha^2}),
\end{equation}
where $r_{s}$ is given by (\ref{radius of the sphere with beta, sigma}), provided  (\ref{alpha min for the general class}) holds. Therefore, all configurations with $\frac{2\, \sqrt{\sigma}}{1 + \sigma} < \alpha < 1$ are inside the inner horizon. In fact, from (\ref{radius of the sphere with beta, sigma}) we get

\begin{equation}
\label{rs/rH for alpha < 1}
\frac{(1 + \sigma)\, (1 + \sqrt{1 - \alpha^2})}{2\, \sigma} < \left(\frac{r_{s}}{r_{-}}\right) < 1, 
\end{equation}
This inequality for $\alpha = 1$  reduces to (\ref{rs/rH for alpha = 1}), as expected.
For a fixed $\sigma$, it narrows as $\alpha$ approaches its minimum (\ref{alpha min for the general class}) in such a way that   $(r_{-} - r_{s}) \rightarrow 0^{+}$, with $r_{-} \rightarrow  \frac{2\, M}{\sigma + 1}$, when $\alpha \rightarrow \tilde{\alpha}_{min}$ (in this limit $r_{+} \rightarrow \sigma\, r_{-}$).

$\bullet$ {\it Central density:}
When  $\sigma = 3$  the central density  is finite;  using (\ref{K in terms of rs}), (\ref{radius of the sphere with beta, sigma}) and (\ref{m of the sphere with beta, sigma}) it can be expressed as 
\begin{equation}
\label{rho M squared for bounded distribution}
\rho_{c} = \frac{81\, \left(3 + \beta\right)^{(3 + \beta)/\beta}\, \left(\beta - 1\right)^{3 (\beta - 1)/\beta}}{128\, \pi\, \alpha^6\, \beta^4 \, M^2}, 
\end{equation}
which reduces to (\ref{rho M squared for unbounded distribution}) for $\beta = 1$. For any given $M$ and $Q$ this expression gives the range of values for $\rho_{c}$ allowed by  $\beta$, which for $\alpha \in (\tilde{\alpha}_{min}, \, 1 ]$ is determined by

\begin{equation}
\label{beta tilde}
\tilde{\beta} < \beta < \infty, \;\;\;\;\;\tilde{\beta} \equiv \frac{3\, (1 + 2\, \sqrt{1 - \alpha^2})}{4\, \alpha^2 - 3}.
\end{equation}
From the last two expressions, we get

\begin{equation}
\label{final expression, inequality for central density}
1 < \frac{128\, \pi\, \rho_{c}\, M^2\, \alpha^6}{81} < \frac{1}{\tilde{\beta}^4}\,{\left[\left(3 + \tilde{\beta}\right)^{(3 + \tilde{\beta})}\, \left(\tilde{\beta} - 1\right)^{3\,(\tilde{\beta} - 1)}\right]^{1/\tilde{\beta}}}.
\end{equation}
To obtain an order of magnitude for $\rho_{c}$ we take $M = M_{\odot}$. Then, from (\ref{final expression, inequality for central density}),  we get
\begin{eqnarray*}
\alpha = 1.00:\;\;\;&&\rho_{c} \in \left(1.25, \, 2.22\right) \times 10^{20} \, kg/m^3,\\
\alpha = 0.99:\;\;\;&&\rho_{c} \in \left(1.32, \, 1.80 \right) \times 10^{20} \, kg/m^3,\\
\alpha = 0.90:\;\;\;&&\rho_{c} \in \left(2.34, \, 2.46\right) \times 10^{20} \, kg/m^3, \\
\alpha = 0.87:\;\;\;&&\rho_{c} \in \left(2.88, \, 2.88\right) \times 10^{20} \, kg/m^3.
\end{eqnarray*}
Thus, the central density is of the order of $10^{20}\, kg/m^3$ in the whole range of $\alpha$. We recall that $\sigma = 3$ requires $\tilde{\alpha}_{min} = \sqrt{3}/2$ so that $\alpha \in (\sqrt{3}/2, \, 1)$, which incidentally saturates the inequality (\ref{values of alpha for which rs less than r-}) obtained on general grounds.

\medskip

$\bullet$ {\it{Tolman-Whittaker mass:}} In the present case the TW mass (\ref{simplified form of TW mass}) inside the spheres can be expressed as

\begin{equation}
\label{TW mass for the case with beta, sigma}
M(x) = \frac{M\, x^{\sigma}\, [(1 + \sigma)\, x^{\beta} - (\sigma - 1)\,(\beta - 1)]}{\beta\,(1 + \sigma)}\, \left[\frac{\sigma + \beta}{(1 + \sigma)\, x^{\beta} + \beta - 1}\right]^{(\sigma + \beta)/\beta}.
\end{equation}
Evaluating this at the boundary surface ($x = 1$) we get

\begin{equation}
\label{TW mass for the case with beta, sigma at the surface}
M(r_{s}) = - \frac{M\, (\sigma - 1)}{\beta\, (1 + \sigma)}\, \left[\beta - \frac{2\, \sigma}{\sigma - 1}\right].
\end{equation}
As expected, a sphere with $\alpha \leq 1$ (for which $\beta > \frac{2\, \sigma}{\sigma - 1}$)  is gravitationally repulsive not only in its interior  but also and its vicinity, although it is covered by an horizon.

If $\alpha  > 1$, then the vicinity of the spheres can be gravitationally attractive or repulsive,  
depending{\textemdash}respectively{\textemdash}on whether $\beta$ is less or greater than $2\,\sigma/(\sigma - 1)$. This can be detected by an external observer since there are no horizons.

\subsection{Dymnikova's nonsingular black hole}

Next,  we consider Dymnikova's spacetime \cite{Dymnikova1}, which{\textemdash}in our notation{\textemdash}is generated by the function

\begin{equation}
\label{Dymnikova's solution}
f(r) = 1 - \frac{a\, b}{r}\, \left(1 - \mbox{e}^{- r^3/a^3}\right),
\end{equation}
where $a$ and $b$ are  positive constants; $a$ has dimensions of length and $b$ is dimensionless. This line element has widely been discussed in the literature with about $272$ citations (adsabs.harvard.edu/abs/1992GReGr..24..235D); it is de Sitter-like near the center and resembles the Schwarschild metric with total mass  $= (a \,b)/2$ at large distances ($r \gg a$). The presence of horizons is governed by the parameter $b$. Namely, (\ref{Dymnikova's solution}) has no zeros if $b < b_{crit} \approx 1.456$, one double zero at $r \approx 1.235\, a$ if $b = b_{crit}$, and two simple zeros if $b > b_{crit}$. It can be interpreted as an exact solution of the Einstein equations coupled to nonlinear electrodynamics describing a magnetic monopole.\footnote{As a matter of fact, from \cite{FanWang} it follows that  the Einstein field equations (\ref{equations for gravity}) evaluated   for the metric (\ref{metric in temrs of f}) are automatically satisfied for any  function $f(r)$ if   the EMT given by (\ref{explicit form of F}), (\ref{T00 = T11, general case}) and (\ref{T22 = T33, general case}) describes  a  magnetic dipole, i.e., when $E(r) = 0$ and $B(r) = Q_{m}/r^2$, where $Q_{m}$ is a constant  coming from the integration of the Bianchi identities (\ref{Bianchi identity fro the EM field}). In which case the function $L(F)$, where $F = 2\, B^2$, is given by $L = 2\, G_{0}^{0}$ with $r = \left(\frac{2\, Q_{m}^2}{F}\right)^{1/4}$. For Dymnikova's spacetime (\ref{Dymnikova's solution}) one can easily find $L(F) = \frac{6\, b}{a^2}\,\exp{\left[- \left(\frac{2\, Q_{m}^2}{a^4\, F}\right)^{3/4}\right]}$.}

\medskip

Below we will see that Dymnikova's spacetime can be used to describe the interior of regular charged spheres, whose properties are essentially the same as those considered in the preceding subsection. 

\medskip

The effective EMT supporting  (\ref{Dymnikova's solution}) is
\begin{eqnarray}
\epsilon = - p_{r} &=& \frac{3\, b}{8\, \pi\, a^2}\,\mbox{e}^{- r^3/a^3},\label{density and radial pressure} \\
p_{\perp} &=& \epsilon\, \left(- 1 + \frac{3\, r^3}{2\, a^3}\right).\label{tangential pressure}
\end{eqnarray}
We note that  the DEC ($\epsilon \geq |p_{\perp}|$) is only satisfied in the interior region
where $\left(\frac{r}{a}\right) \leq \left(\frac{4}{3}\right)^{1/3}$. If we  restrict the use of (\ref{Dymnikova's solution}) to that region, then the matter content that generates (\ref{Dymnikova's solution}) can be interpreted as a charged perfect fluid, whose energy density $\rho$, pressure $p$ and electric field intensity $E$ are given by

\begin{eqnarray}
\rho(r) = - p(r) &=& \frac{3\, b}{32\, \pi\, a^2}\,\left(4 - \frac{3\, r^3}{a^3}\right)\label{rho for Dymnikovas solution}, \\
E^2(r) &=& \frac{9\, b\, r^3}{4\, a^5}\,\mbox{e}^{- r^3/a^3}\label{E inside the sphere},
\end{eqnarray}
which have been obtained by substituting (\ref{density and radial pressure}){\textendash}(\ref{tangential pressure}) into (\ref{rho, general expression}) and (\ref{E, general expression}).

The energy density $\rho(r)$ is positive and drops continuously from its maximum value at the center to zero at the surface $r_{s} = \left(\frac{4}{3}\right)^{1/3} a$. Therefore, $r_{s}$ represents the outer boundary of the charged fluid sphere, i.e.,

\begin{equation}
\label{a in terms of rs}
a = \left(\frac{3}{4}\right)^{1/3}\, r_{s}.
\end{equation}
The boundary conditions (\ref{BC 1}) and (\ref{BC 2}) provide two equations from which we get $b$ and $r_{s}$,

\begin{eqnarray}
b &=& \frac{16 \times 6^{2/3} \times\mbox{e}^{4/3} }{3\, \alpha^2 \, (3 + \mbox{e}^{4/3})^2}, \label{solution for b}\\
r_{s} &=& \frac{M\, \alpha^2}{8}\, (3 + \mbox{e}^{4/3}).\label{rs}
\end{eqnarray}
In this parametrization the original function (\ref{Dymnikova's solution}) becomes

\begin{equation}
\label{f(x) for Dymnikova's spacetime}
f(x) = 1 - \frac{16\, \mbox{e}^{4/3}}{\alpha^2\, x\, (3 + \mbox{e}^{4/3})^2}\, \left(1 - \mbox{e}^{- 4\, x^3/3}\right), \;\;\;\;x = \frac{r}{r_{s}}, \;\;\; 0 \leq x \leq 1.
\end{equation}
As in the previous case, the charge-to-mass parameter $\alpha$ is bounded below by the condition  $f(x) > 0$. Since (\ref{f(x) for Dymnikova's spacetime}) is a decreasing function of $x$, this condition is fulfilled if $f(1) > 0$, which in turn requires

\begin{equation}
\label{restriction on alpha}
\alpha > \bar{\alpha}_{min} = \frac{4\, \sqrt{\mbox{e}^{4/3} - 1}}{(3 + \mbox{e}^{4/3})} \approx 0.98.
\end{equation}
It is easy to verify that this inequality is the solution to (\ref{rs less than inner horizon}) with $r_{s}$ given by (\ref{rs}). 

$\bullet$ Consequently,
 when $\bar{\alpha}_{min} < \alpha \leq 1$ the charged spheres are located inside the horizon.
In fact, when $\alpha = 1$ from (\ref{rs}) we obtain

\[
\left(\frac{r_{s}}{r_{\ast}}\right) = \frac{3 + \mbox{e}^{4/3}}{8} \approx 0.85.
\]
Similarly,  when  $\bar{\alpha}_{min} < \alpha < 1$  we get

\[
\left(\frac{r_{s}}{r_{-}}\right) = \frac{3 + \mbox{e}^{4/3}}{8}\, (1 + \sqrt{1 - \alpha^2}) \approx 0.85 \, (1 + \sqrt{1 - \alpha^2}),
\]
which is  approximately $0.85 <\left(r_{s}/r_{-}\right) < 1$ for $\alpha \in (\bar{\alpha}_{min}, \, 1)$.

\medskip

$\bullet$ Regarding the central density  $\rho_{c}$; from (\ref{rho for Dymnikovas solution}), (\ref{a in terms of rs}), (\ref{solution for b}) and (\ref{rs}) we find 

\begin{equation}
\label{central density for Dymnikova's charged spheres}
\rho_{c} = \frac{2^{9}\, \mbox{e}^{4/3}}{\pi\, (3 + \mbox{e}^{4/3})^4 \, \alpha^6\, M^2}. 
\end{equation}
Thus, for $\alpha \in (\bar{\alpha}_{min}, \, 1)$ we get $0.29 < \rho_{c}\, M^2 < 0.32$, approximately. Consequently, for a body of $M =  M_{\odot}$ the central density is about $\rho_{c} = (1.79{\textendash}1.98) \times 10^{20}\, kg/m^3$.

\medskip

$\bullet$ The TW mass inside the spheres is given by

\begin{equation}
\label{TW mass for Dymnikova's charged spheres}
M(x) =   \frac{M}{1 + 3\, \mbox{e}^{- 4/3}} \,      \left[1 - (1 + 4\, x^3) \,\mbox{e}^{(- 4 x^3/3)}\right],
\end{equation}
where we have used (\ref{simplified form of TW mass}), (\ref{rs}),  and (\ref{f(x) for Dymnikova's spacetime}). 
On the other hand, from (\ref{TW mass for RN}) and (\ref{rs}) it follows that $M_{\!_{RN}}(r)$ is negative  for $r < \left(\frac{8  \, r_{s}}{3 + \mbox{e}^{4/3}}\right) \approx 1.18 \, r_{s} $. Consequently, unlike the previous case, all these charged spheres  possess  negative TW mass not only inside but also in their neighborhood, regardless of the choice of $\alpha$. When $\alpha \leq 1$ the region of negative TW mass is hidden behind the horizon, but not when $\alpha > 1$ because there are no horizons.

\section{Final comments and summary}

The popular belief is that at the core of a BH there is a singularity covered by an event horizon.    This notion is a cumulative result of various far-reaching developments,  among which the more influential are (a) the understanding that the gravitational collapse of a homogeneous spherical dust cloud{\textemdash}as it evolves in time{\textemdash}leads to the formation of a singularity covered by an event horizon \cite{Oppenheimer}; (b) the singularity theorems of Geroch, Hawking,
and Penrose \cite{Hawking} which{\textemdash}in the framework of general relativity and the assumption that certain general conditions hold{\textemdash}prove that a sufficiently massive collapsing object will undergo continual gravitational collapse,
resulting in the formation of a gravitational singularity; (c) the (weak) cosmic censorship hypothesis which asserts that these singularities are always hidden inside a BH (see e.g., \cite{Hod} and references therein).

However, by definition a BH is a region of an asymptotically  flat spacetime from which it is impossible to send signals to infinity. Therefore,  its characterizing feature is the appearance of an event horizon, not the presence (or absence) of a spacetime singularity. 
As mentioned in the Introduction, there are several solutions of Einstein's field equations{\textemdash}with different sources{\textemdash}which have horizons but no spacetime singularities. 
Most of them are  represented by 
 static, spherically symmetric metrics of the form (\ref{metric in temrs of f}) which are (i) regular everywhere, (ii) asymptotically flat, and (iii) $f(r)$ possesses a minimum, say $f_{min}$, such that, for $r > 0$, it has no zeros if $f_{min} > 0$, two simple zeros if $f_{min} < 0$ and one double zero if $f_{min} = 0$.  
When such spacetimes are used in the whole range $0 \leq r < \infty$ these three cases describe, respectively,  a regular spacetime, a regular non-extreme BH with both outer and inner Killing horizons, and a regular extreme BH with degenerate Killing horizon.

If the Einstein equations (\ref{equations for gravity}) are used to evaluate the effective EMT, these  regular BHs are supported by anisotropic matter with finite energy density and pressures which have a de Sitter-like behavior in the central region, as envisaged by Gliner \cite{Gliner}. Therefore, in that region the components of the EMT obey the dominant but not the strong energy condition. Some of these spacetimes have been interpreted in terms of electric or magnetic monopoles in a suitable chosen nonlinear electrodynamics.

In this work we have shown that these regular BH  metrics can also be interpreted as exact solutions of the Einstein equations coupled to ordinary linear electromagnetism{\textemdash}i.e., as sources of the RN spacetime. We have constructed regular RN black holes,  where  the central singularity is replaced by a regular   perfect fluid charged sphere which, for the case where $|Q| < M$ ($Q = M$), is located inside the inner (Killing) horizon. The coordinate radius of the sphere is determined by the equation $\epsilon =  p_{\perp}$ and is expressed in terms of $M$, $Q$  and the parameters of the solutions. It is  important  to emphasize that the condition  $f > 0$ is fulfilled if and only if  the spheres are inside  the inner horizon (Killing horizon  when $|Q| = M$). This condition also imposes a lower bound on the charge-to-mass parameter $\alpha$. In Newtonian terms, it provides the electrostatic repulsive force, required by the TOV equation (\ref{TOV}),  to balance the inward hydrostatic force produced by the gradient of the negative pressure; if $\alpha$ were less than the required minimum,   then $f$ would be negative in the outermost layers   and the perfect fluid sphere could not be in static equilibrium.
   Regarding the central mass density, we have seen that it  is practically insensitive to the concrete value of $\alpha$; for an object of one solar mass it  is about $10^{20}\, kg/m^3$, which is thousands times greater than the approximate density of an atomic nucleus. However, an increase (decrease) of $M$ leads to a decrease (increase) of this quantity.

It should be mentioned that in the present interpretation there is no room for linear (Einsten-Maxwell) magnetic monopoles, even if the original spacetime (\ref{metric in temrs of f}) is attributed to a magnetic monopole in a nonlinear electromagnetic theory. Also, the lower limit on $\alpha$ heavily depends on the model. For example, the general BH metrics (\ref{general function}){\textemdash}as well as the thin-shell models \cite{Lemos1, Nami}{\textemdash}saturate the requirement  (\ref{values of alpha for which rs less than r-});  meanwhile for the Dymnikova's spacetime 
it is considerably more restricted (\ref{restriction on alpha}), although it can still be less than $1$, allowing the existence of static charged spheres inside the RN horizons. For the   Beato and Garcia solution  \cite{{Beato2}} (not discussed here), $\alpha > 1$.

The inner horizon in the standard RN solution   is unstable under ``small" external perturbations \cite{Chandrasekhar}. The full nonlinear nature of this instability{\textemdash}dubbed ``mass-inflation"{\textemdash}was discovered  by 
Poisson and Israel \cite{Poisson1} and has been confirmed   in a number of scenarios (see, e.g.,  \cite{Droz, Dafermos, Pedro}).
On the other hand, Dymnikova and Galaktionov  \cite{Dymnikova4} demonstrated  that any configuration described by a spherically symmetric geometry with a de Sitter center is stable
to axial perturbations, and{\textemdash}in the case
of the polar perturbations{\textemdash}they found the criteria for stability of Dymnikova's nonsingular black hole \cite{Dymnikova1}.  At first glance, both results do not appear to be mutually consistent. Therefore, given that the  predicted charged spheres  are located inside the inner RN horizon and are de Sitter at the center, the next step of this research  would be to  study their stability.
However, that is beyond  the scope of the present work.

\end{document}